\documentclass[10pt,a4paper,twocolumn]{article}

\usepackage[utf8]{inputenc}
\usepackage[T1]{fontenc}
\usepackage[margin=20mm,columnsep=7mm]{geometry}
\usepackage{graphicx}
\usepackage{amsmath,amssymb,amsfonts}
\usepackage{textcomp}
\usepackage{xcolor}
\usepackage{caption}
\usepackage{booktabs}
\usepackage{authblk}
\usepackage[numbers,sort&compress]{natbib}
\usepackage[hidelinks]{hyperref}
\usepackage{microtype}

\newcommand{\tNi}{t_{\mathrm{Ni}}}

\newcommand{\Meff}{M_{\mathrm{eff}}}
\newcommand{\Ks}{K_{\mathrm{s}}}
\newcommand{\Kv}{K_{\mathrm{v}}}
\newcommand{\Ms}{M_{\mathrm{s}}}

\title{Magnetic Anisotropy and Magnetization Dynamics in Si/Ni Multilayers Probed by FMR and VSM Magnetometry}

\author{Dushyant Singh}
\author{Subha Maity}
\author{Tharundev V V}
\author{Subhavenkadesh SS}
\author{Vigneshwari SR}
\author{Krista R. Khiangte$^{*}$}
\affil{Department of Physics, Indian Institute of Technology Gandhinagar, Palaj, Gujarat 382055, India\\
\normalsize $^{*}$Corresponding author: \href{mailto:krista.khiangte@iitgn.ac.in}{krista.khiangte@iitgn.ac.in}}
\date{}

\begin{document}

\twocolumn[{%
\begin{@twocolumnfalse}
\maketitle

\begin{abstract}
The magnetic anisotropy and magnetization dynamics of sputtered [Si(40~\AA)/Ni($t_{\mathrm{Ni}}$~\AA)]$_{20}$ multilayers were systematically investigated at room temperature using SQUID--VSM magnetometry and angle-dependent ferromagnetic resonance (FMR). Static measurements reveal soft ferromagnetic behaviour in thicker Ni layers and mixed ferromagnetic and superparamagnetic contributions at reduced thicknesses. The saturation magnetization increases toward the bulk Ni value with increasing thickness, consistent with Ni nanocrystallite growth. Magnetic dead-layer analysis yields an interfacial dead-layer thickness of approximately $9$~\AA\ per interface, accounting for the suppressed magnetization in thinner Ni layers. Angle-dependent FMR measurements reveal effective easy-plane magnetic anisotropy, while out-of-plane spectra exhibit additional non-uniform spin-wave modes. The effective magnetization remains lower than the static saturation magnetization, indicating a perpendicular anisotropy contribution that partially compensates the demagnetizing field. Quantitative anisotropy analysis yields an interfacial anisotropy constant of $K_S = +0.17~\mathrm{erg\,cm^{-2}}$, favouring perpendicular magnetization, and a volume anisotropy constant of $K_V = -1.23 \times 10^6~\mathrm{erg\,cm^{-3}}$, favouring in-plane alignment, with an extrapolated anisotropy crossover thickness of $t_c \approx 27$~\AA. Increasing FMR line-shape asymmetry with decreasing Ni thickness is consistent with enhanced interfacial roughness. These findings highlight the competing interfacial and volume anisotropies and the importance of Ni thickness and interface quality in tailoring Si/Ni multilayers for silicon-compatible spintronic applications.

\end{abstract}
\vspace{1.5em}
\end{@twocolumnfalse}
}]

\section{Introduction}\label{sec1}

Magnetic multilayers (MLs) are widely investigated because of their interesting magnetic properties and possible applications in magnetic sensors, recording media, and spintronic devices~\cite{Zhang2008SSC, Leitao2024, Ikeda2010}. These systems consist of alternating magnetic and nonmagnetic layers, whose properties can be controlled by choosing suitable constituent materials and layer thicknesses. The introduction of semiconducting spacer layers, such as Si, Ge, GeSn, GaAs, and GaN, provides an additional means of modifying the magnetic behaviour through interface effects~\cite{Liu2007JAP, Lucinski2011OP, Grundy2000PRB, Haque2001PB, Tharundev2026}. Consequently, such structures offer opportunities for combining magnetic functionalities with semiconductor technology. Among these systems, Si/Ni structures are of particular interest owing to their relevance to silicon based heterostructures, complementary metal oxide semiconductor (CMOS) technology, field-effect transistors, Schottky diodes, and optical recording applications~\cite{Ashery2010ME,Breil2015ME,Saha2006SSE,Ou2014TSF,Chang2019SR}.

One of the important properties governing the magnetic behaviour of MLs is magnetic anisotropy, which describes the preference of the magnetization to align along a particular direction~\cite{Daalderop1992PRL}. In thin-film systems, this preferred direction depends on the magnetic layer thickness, surface morphology, and interfacial characteristics. With decreasing magnetic layer thickness, the contribution of surfaces and interfaces becomes increasingly significant compared with that of the magnetic volume \cite{singh2025viable}. Under suitable conditions, these contributions favour the alignment of magnetization normal to the film plane, giving rise to perpendicular magnetic anisotropy (PMA)~\cite{Johnson1996RPPa,SenthilKumar2003PB,Svedberg2002JAP,LeeHone2017JMMM,Gopman2016SR}. The study of PMA is therefore important for developing high-density magnetic recording devices. In addition, materials exhibiting PMA are useful for spin transfer torque magnetic random access memory (STT-MRAM), where suitable anisotropy can support thermal stability, lower switching energy, and increased storage density~\cite{Tudu2017Sensors}.

Several investigations have been carried out to understand magnetic anisotropy in Ni/NiO, Ni/Co, Ni/Cu, Ni/Pd, Ni/Fe, Co/Pd, and Fe-based MLs ~\cite{Poulopoulos2010APL,Arora2017PRB,Lassri2001TSF,Chafai2011JMMM,Kyuno1996JJAP,Tudu2017Sensors,SenthilKumar2003PB}. These studies indicate that the interfacial anisotropy depends on both the materials forming the MLs and the nature of their interfaces. For example, positive interfacial anisotropy contributions have been reported in Ni/Cu and Ni/Pd MLs, whereas negative contributions occur in other Ni-based structures~\cite{Johnson1996RPPa,Lassri2001TSF,Chafai2011JMMM}. Furthermore, roughness, intermixing, and structural defects modify the local environment of magnetic atoms and can consequently affect the anisotropy. It is therefore necessary to examine the structural and microstructural characteristics when discussing the magnetic response of these systems.

The magnetic anisotropy energy (MAE) of thin films and MLs can be studied through static and dynamic magnetization processes~\cite{Johnson1996RPPa}. Static measurements are commonly performed using superconducting quantum interference device (SQUID) magnetometry and vibrating sample magnetometry, whereas ferromagnetic resonance (FMR) and Brillouin light scattering provide dynamic approaches~\cite{Johnson1996RPPa,Bergholz1984JMMM,Purcell1992JMMM,Zijlstra1967}. In particular, magnetization measurements and FMR spectroscopy provide complementary information on the magnetic parameters. Their combined analysis allows the surface and volume anisotropy contributions to be evaluated and their dependence on magnetic layer thickness to be examined.

Although magnetic anisotropy has been investigated in several Ni-based MLs, further study is required to establish how the MAE of Si/Ni MLs changes with Ni thickness and the associated structural characteristics. Our previous work examined the structural and microstructural properties of these MLs~\cite{singh2025viable}, providing the basis for the present investigation. Here, the magnetic anisotropy of Si/Ni MLs is systematically studied at room temperature using SQUID--VSM measurements and FMR spectroscopy. The magnetic parameters obtained through these static and dynamic processes are correlated with the previously reported structural results to understand the influence of Ni layer thickness and interfacial characteristics on the MAE.

\section{Experimental details}

The Si/Ni MLs were fabricated by DC magnetron sputtering onto Si(111) substrates at room temperature. The deposition chamber was evacuated to a base pressure of approximately $1.5\times10^{-6}$~mbar before introducing high-purity argon as the sputtering gas. The working pressure was maintained at $4.0\times10^{-3}$~mbar during deposition, with a sputtering power of 40~W applied to each target.

Prior to deposition, the Si(111) substrates were cleaned ultrasonically using a sequential isopropyl alcohol (IPA)--acetone--IPA procedure for approximately 10~min to remove surface contaminants. The substrates were subsequently etched in a 2\% HF solution for 15~seconds to eliminate the native oxide layer, rinsed thoroughly with deionized (DI) water, and dried under high-purity nitrogen ($\mathrm{N}_2$). The cleaned substrates were promptly transferred into the sputtering chamber, followed by evacuation to the required base pressure and three argon-purging cycles to minimize residual contamination.

The MLs were deposited with a fixed nominal Si thickness of $40\,\text{\AA}$, while the Ni layer thickness ($t_{\mathrm{Ni}}$) was varied to investigate its influence on the magnetic properties. The resulting structures can be represented as $[\mathrm{Si}(40\,\text{\AA})/\mathrm{Ni}(t_{\mathrm{Ni}}\,\text{\AA})]_{20}$, where the subscript denotes $10$, $20$, repetitions of the Si/Ni bilayer. Seven samples with Ni thicknesses of $20$, $30$, $40$, $50$, $60$, $90$, and $130\,\text{\AA}$ were prepared. Further details regarding the deposition procedure, the rationale for selecting a Si layer thickness of approximately $40\,\text{\AA}$, and the structural characterization of the MLs are provided in our previous work~\cite{singh2025viable}.

Magnetic hysteresis measurements were performed at room temperature using a vibrating sample magnetometer (VSM) integrated into a Physical Property Measurement System (PPMS; Quantum Design, USA). Dynamic magnetic properties were investigated using a JES-FA200 electron spin resonance spectrometer (JEOL) operating at $9.443$~GHz. FMR spectra were acquired at room temperature for different orientations of the applied magnetic field relative to the MLs plane to examine the angular dependence of the resonance behavior and magnetic anisotropy.

\section{Results and Discussion}\label{sec3}

\subsection{Structural and microstructural Study}\label{subsec2}

The structural and interfacial properties of the $[\mathrm{Si}(40\,\text{\AA})/\mathrm{Ni}(t_{\mathrm{Ni}}\,\text{\AA})]_{20}$ MLs were examined in the previous study~\cite{singh2025viable}. The GIXRD patterns showed a prominent Ni (111) reflection, whereas the Ni (200) and Ni (220) peaks were observed for $t_{\mathrm{Ni}}\geq40\,\text{\AA}$. These observations indicate the polycrystalline nature of the fcc-Ni layers, with preferential growth along the (111) plane. Furthermore, the average crystallite size estimated using the Scherrer formula decreased from approximately $68\,\text{\AA}$ to $15\,\text{\AA}$ as the Ni layer thickness was reduced from $130\,\text{\AA}$ to $20\,\text{\AA}$. The appearance of modulation Bragg peaks in the XRR patterns confirmed the periodic continuous MLs structure. In addition, the decrease in the relative intensity of higher-order Bragg peaks with decreasing $t_{\mathrm{Ni}}$ indicated an increase in Si/Ni interface roughness. HRTEM images and SAED patterns further revealed crystalline Ni grains within the Ni layers and the amorphous nature of the Si layers. At larger Ni thicknesses, the grains were interconnected, whereas reducing $t_{\mathrm{Ni}}$ resulted in smaller, more isolated crystallites with less distinct grain boundaries. Such thickness-dependent structural changes increase the relative contribution of interfaces and are expected to influence intergranular transport and the magnetic behaviour of the MLs stack.

\subsection{Magnetic anisotrpy study}

\begin{figure}[!tbp]
    \centering
    \includegraphics[width=\linewidth]{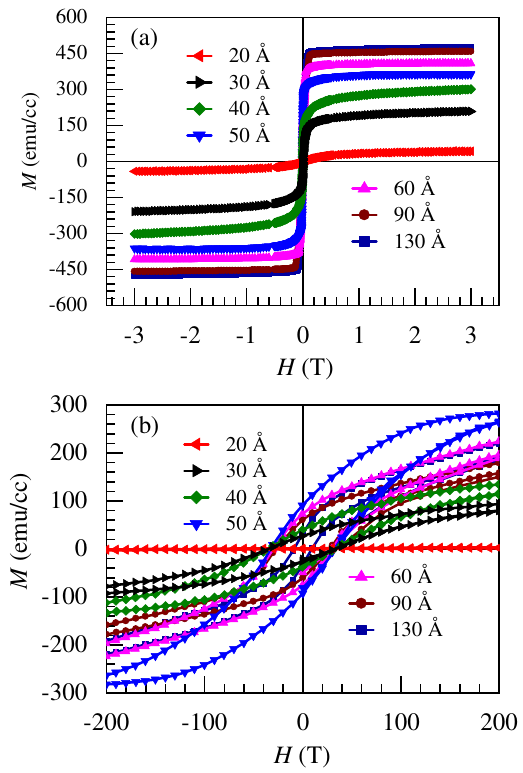}
    \caption{(a) In-plane $M$--$H$ hysteresis loops measured up to an applied magnetic
field of $20\,\mathrm{kOe}$ for the
$[\mathrm{Si}(40\,\text{\AA})/\mathrm{Ni}(t_{\mathrm{Ni}}\,\text{\AA})]_{20}$
MLs at room temperature. (b) Enlarged views of the low-field regions
of the hysteresis loops presented in panel (a), highlighting the determination
of the coercive field, $H_c$. In panels (b), the original symbols and colour
scheme are retained for consistency with panel (a).}
    \label{fig:MH}
\end{figure}

To investigate the influence of Ni layer thickness on the magnetic response, in-plane magnetic field-dependent magnetization (M(H)) loops were measured at 300~K with a maximum applied field of 3~T. The resulting $M$--$H$ curves of the $[\mathrm{Si}(40\,\text{\AA})/\mathrm{Ni}(t_{\mathrm{Ni}}\,\text{\AA})]_{20}$ MLs are presented in Fig.~\ref{fig:MH}(a). The samples exhibit different approaches to saturation depending on $t_{\mathrm{Ni}}$. For $50\,\text{\AA}\leq t_{\mathrm{Ni}}<130\,\text{\AA}$, a comparatively small field is sufficient to bring the magnetization close to saturation, which is characteristic of soft ferromagnetic behaviour. On reducing the Ni thickness to $20\,\text{\AA}\leq t_{\mathrm{Ni}}\leq40\,\text{\AA}$, the magnetization continues to increase over a wider field range before approaching saturation. Such a response is consistent with the presence of ferromagnetic (FM) and superparamagnetic (SPM) contributions, as suggested by the previously reported zero-field-cooled (ZFC) and field-cooled (FC) measurements on these MLs~\cite{singh2025viable} and other systems \cite{Singh2024skew, VSR2026Co, VSR2026Fe}. The successive changes in the hysteresis curves further suggest a gradual evolution between these magnetic regimes. An enlarged view of the low-field regime is provided in Fig.~\ref{fig:MH}(b) to compare the coercivity ($H_{\mathrm{c}}$), retaining the sample symbols and colours used in Fig.~\ref{fig:MH}(a).

\begin{figure}[!tbp]
    \centering
    \includegraphics[width=\linewidth]{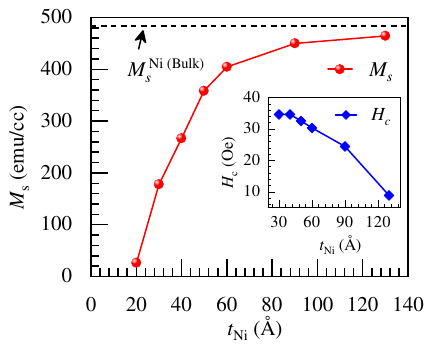}
    \caption{The saturation magnetization, $M_s$, extracted from the $M$--$H$ hysteresis
loops in Fig.~\ref{fig:MH}(a) as a function of the Ni layer thickness,
$t_{\mathrm{Ni}}$. The horizontal dashed line indicates the bulk saturation
magnetization of Ni at $300\,\mathrm{K}$, approximately
$484\,\mathrm{emu\,cm^{-3}}$. The inset shows the variation of the coercive
field, $H_c$, with $t_{\mathrm{Ni}}$ for the MLs obtained from Fig.~\ref{fig:MH}(b).}
    \label{fig:Ms}
\end{figure}

Fig.~\ref{fig:Ms} shows the variation of the saturation magnetization
($M_{\mathrm{s}}$) with Ni layer thickness, $t_{\mathrm{Ni}}$. The
$M_{\mathrm{s}}$ values were extracted from the $M(H)$ loops shown in
Fig.~\ref{fig:MH}(a) by extrapolating the high-field magnetization data
to zero field, following the procedure described in Ref.~\cite{Singh2022SD}.

A substantial thickness dependence is observed for the thinner Ni layers, followed by a weaker variation as the layers become thicker. With increasing $t_{\mathrm{Ni}}$, the magnetization tends towards the bulk Ni value of approximately $484\,\mathrm{emu\,cm^{-3}}$~\cite{Cullity2011, Singh2020JMMM, Singh2022SD}. As established in our previous structural investigation, the average crystallite size ($D$) associated with the Ni (111) reflection increases with increasing Ni layer thickness, $t_{\mathrm{Ni}}$ (see the fourth
column of Table~1 in Ref.~\cite{singh2025viable}). The simultaneous
increase in $D$ and $M_{\mathrm{s}}$ suggests that the growth of Ni
nanocrystallites contributes to the evolution of the magnetic response
toward its bulk-like behavior.

The inset of Fig.~\ref{fig:Ms} shows the coercivity as a function of $t_{\mathrm{Ni}}$. Unlike $M_{\mathrm{s}}$, $H_{\mathrm{c}}$ decreases as the Ni layers become thicker, reflecting easier magnetization reversal. A coercivity of approximately $34\,\mathrm{Oe}$ is obtained for the sample with $t_{\mathrm{Ni}}=20\,\text{\AA}$. The enhanced coercivity at smaller thicknesses may arise from the combined influence of the finer grain structure, interfacial anisotropy, and weakened magnetic interactions between neighboring grains (see the ZFC-FC discussion of Fig. 5 in Ref.~\cite{singh2025viable}). Although an SPM contribution may be present in these samples, it does not by itself account for the finite hysteresis.

\begin{figure}[!tbp]
    \centering
    \includegraphics[width=\linewidth]{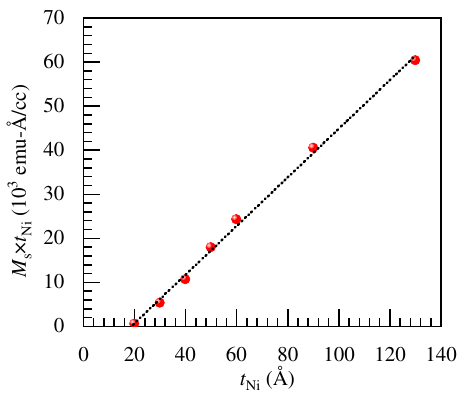}
    \caption{The product of the saturation magnetization and Ni layer thickness,
$M_s.t_{\mathrm{Ni}}$, plotted as a function of $t_{\mathrm{Ni}}$.
Solid red sphere symbols represent the experimental measurements, while the dotted
curves correspond to fits based on Eq.~\eqref{eq:dead_layer}.}
    \label{fig:deadlayer}
\end{figure}

The magnetization measurements also allow the thickness of the magnetically inactive region near the Si/Ni interfaces to be estimated. The formation of a magnetic dead layer at the Si/Ni interfaces reduces the effective magnetic thickness of the Ni layers and is therefore important for understanding the suppression of magnetization in these MLs~\cite{ElKhiraoui2008}. This region, commonly referred to as the magnetic dead layer, reduces the effective magnetic thickness of the Ni layer. To examine its contribution, the thickness dependence of $M_{\mathrm{s}}$ was analysed using the conventional dead-layer model~\cite{Singh2022SD},

\begin{equation}
M_{\mathrm{s}}(t_{\mathrm{Ni}})
= M_{0}\left(1-\frac{2\delta}{t_{\mathrm{Ni}}}\right),
\label{eq:dead_layer}
\end{equation}

where $M_{0}$ represents the saturation magnetization of the magnetically active Ni region and $\delta$ denotes the dead-layer thickness at each interface, assuming equal contributions from the two interfaces. Eq.~\ref{eq:dead_layer} can be rearranged as $M_{\mathrm{s}}.t_{\mathrm{Ni}}=M_{0}.t_{\mathrm{Ni}}-2M_{0}.\delta$. Accordingly, the variation of $M_{\mathrm{s}}.t_{\mathrm{Ni}}$ with $t_{\mathrm{Ni}}$ at 300~K is plotted in Fig.~\ref{fig:deadlayer} and fitted with a straight line. The slope gives an extrapolated magnetization of
$M_{0}\approx500\,\mathrm{emu\,cm^{-3}}$, consistent with the value
reported for Ni/V MLs at 300~K by Abid et al.~\cite{Abid1999JMMM}
and comparable to the bulk Ni magnetization of
$484\,\mathrm{emu\,cm^{-3}}$~\cite{Cullity2011}. The intercept gives $\delta\approx9\,\text{\AA}$, slightly exceeding the value reported for discontinuous Si/Ni MLs~\cite{Singh2022SD}. The slightly lower $M_{\mathrm{s}}$ values obtained for the thicker Ni layers compared with the bulk Ni value of approximately $484\,\mathrm{emu\,cm^{-3}}$ may be attributed to magnetically inactive regions of $\delta\approx9\,\text{\AA}$ on both sides of the Ni layer near the Si/Ni interfaces. The total dead-layer thickness ($2\delta = 18\,\text{\AA}$) nearly equals the nominal Ni thickness for $t_{Ni}=20\,\text{\AA}$, leaving only a small magnetically active region, consistent with the observed SPM contribution. For $t_{Ni}\geq20\,\text{\AA}$, the relative contribution of the dead layer decreases, favouring soft ferromagnetic behaviour, as shown in Fig.~\ref{fig:Ms}.

\begin{figure}[!tbp]
    \centering
    \includegraphics[width=\linewidth]{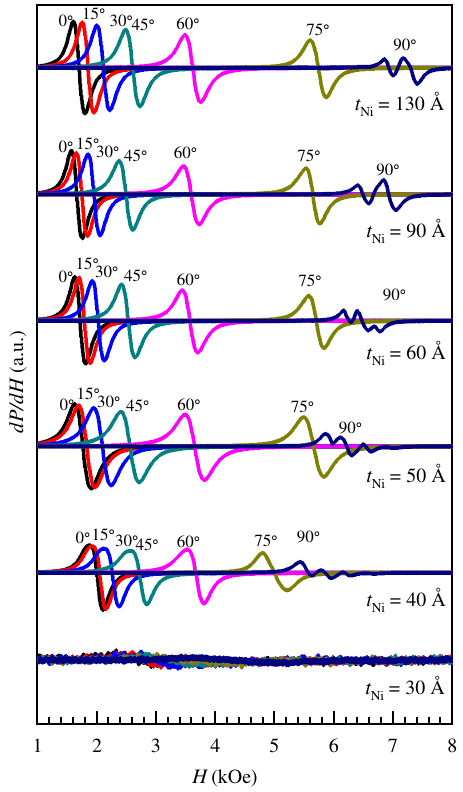}
    \caption{FMR spectra of the
$[\mathrm{Si}(40\,\text{\AA})/\mathrm{Ni}(t_{\mathrm{Ni}}\,\text{\AA})]_{20}$
MLs measured at ambient temperature and a microwave frequency of
$9.443\,\mathrm{GHz}$ for different orientations of the applied magnetic
field. The angle $\theta$ is defined with respect to the sample surface and
varied in increments of $15^\circ$. }
    \label{fig:fmr}
\end{figure}

\begin{figure}[!tbp]
    \centering
    \includegraphics[width=\linewidth]{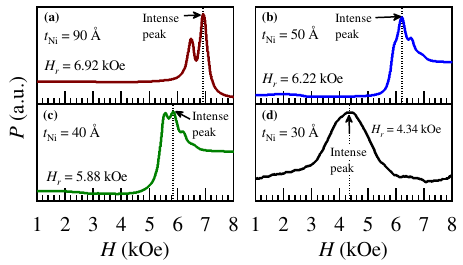}
    \caption{FMR absorption profiles at $\theta = 90^\circ$, obtained by integrating the measured FMR spectra for Si/Ni MLs with Ni thicknesses of (a) $90\,\text{\AA}$, (b) $50\,\text{\AA}$, (c) $40\,\text{\AA}$, and (d) $30\,\text{\AA}$. The dashed vertical lines indicate the resonance fields corresponding to the strongest peaks.}
    \label{fig:FMRint}
\end{figure}

To examine the MAE through, angle-dependent FMR spectra of the $[\mathrm{Si}(40\,\text{\AA})/\mathrm{Ni}(t_{\mathrm{Ni}}\,\text{\AA})]_{20}$ MLs were recorded at 300~K using a microwave frequency of 9.443~GHz. The external magnetic field was rotated from the in-plane ($\theta=0^\circ$) to the perpendicular ($\theta=90^\circ$) orientation in steps of $15^\circ$, where $\theta$ is measured relative to the film surface. Fig.~\ref{fig:fmr} presents the field derivatives of the microwave absorption power, $dP/dH$, for samples with $t_{\mathrm{Ni}}=30$, 40, 50, 60, 90, and 130~\AA{}. Well-resolved resonance signals are observed for $t_{\mathrm{Ni}}\geq40$~\AA{}, whereas the sample with $t_{\mathrm{Ni}}=30$~\AA{} exhibits only a weak response without a clearly distinguishable resonance under the present measurement conditions. No detectable FMR signal was obtained for the samples with $t_{\mathrm{Ni}}=10$ and 20~\AA{}. The absence of a resolved resonance may be associated with superparamagnetic behaviour in the $t_{\mathrm{Ni}}=10$~\AA{} sample and paramagnetic behaviour in the $t_{\mathrm{Ni}}=20$~\AA{} sample near room temperature, as suggested by previously reported ZFC--FC magnetization measurements~\cite{singh2025viable}. For the samples showing pronounced FMR spectra, the dominant resonance shifts progressively towards higher fields as the applied field is rotated away from the film plane. The lowest resonance fields occur near the parallel orientation, while the highest are observed near the perpendicular configuration. This angular variation demonstrates the directional dependence of the magnetic response. Furthermore, the separation between the in-plane and perpendicular resonance fields generally increases with Ni layer thickness.

The spectra recorded at parallel and intermediate orientations are largely dominated by a single derivative resonance. Near the perpendicular orientation, additional features become apparent, particularly for the thicker Ni layers, indicating a more complex mode structure. These features may be associated with nonuniform spin-wave excitations or spatial variations in the effective magnetic field arising from surface/interface roughness~\cite{Lassri2004PB, Singh2022SD}. Surface anisotropy and coupling between neighbouring Ni layers across the Si spacers may also influence the resonance behaviour~\cite{Kordecki1993JMMM,Jin1996PSS}. However, the assignment of individual features to uniform, spin-wave, or surface-localized modes requires further analysis of their resonance fields and angular evolution.

Because the out-of-plane FMR derivative spectra exhibit multiple oscillatory features, identifying the dominant resonance requires careful analysis of the corresponding absorption spectra. Therefore, the derivative spectra were integrated to obtain the absorption profiles, as shown in Fig.~\ref{fig:FMRint} for $t_{\mathrm{Ni}} = 90$, 50, 40, and 30~\AA{} at $\theta = 90^\circ$. The integrated spectra exhibit a dominant absorption peak, marked by a vertical dashed line, alongside weaker features associated with additional magnetic excitations. The dominant peak is assigned to the uniform mode, with resonance field $H_\perp$~\cite{Lassri2001TSF,Maksymowicz1983JMMM}. This mode corresponds to coherent, in-phase precession of the magnetic moments with wavevector $k=0$~\cite{Lassri2001TSF,Lassri2004PB}. Within this interpretation, the weaker resonances on the lower-field side are attributed to nonuniform spin-wave modes with real, nonzero $k$~\cite{Lassri2004PB}. Those on the higher-field side are assigned to surface-localized modes, characterized by an imaginary wavevector $k=ik_{\mathrm{s}}$. Such surface modes have been investigated theoretically by Puszkarski~\cite{Puszkarski1979PSS} and Zhang et al.~\cite{Zhang1987PRB} and observed experimentally in magnetic thin films~\cite{Maksymowicz1983JMMM,Hurdequint1991JMMM}.
The resonance field of the surface-localized mode is expressed as
\begin{equation}
H_{\mathrm{surf}}
=
H_\perp+\frac{2A}{\Ms}k_{\mathrm{s}}^{2},
\label{eq:hsurf}
\end{equation}
where $A$ denotes the exchange stiffness and $k_{\mathrm{s}}$ is the inverse localization length of the surface excitation. For fcc Ni, $A=1.23\times10^{-6}$~erg/cm is adopted~\cite{Vaz2008RPP}. The positive exchange contribution places the surface-mode resonance at a higher field than the uniform mode. Accordingly, the field separation, $H_{\mathrm{surf}}-H_\perp$, can be used to estimate $k_{\mathrm{s}}$. Determination of the surface anisotropy constant ($\Ks$) further requires an appropriate magnetic boundary condition connecting $k_{\mathrm{s}}$ with the interfacial anisotropy.

\begin{figure}[!tbp]
    \centering
    \includegraphics[width=\linewidth]{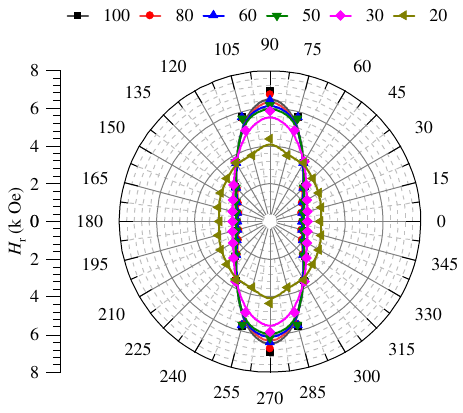}
    \caption{Polar representation of the angular dependence of the ferromagnetic resonance
field, $H_{\mathrm{r}}$, for the
$[\mathrm{Si}(40\,\text{\AA})/\mathrm{Ni}(t_{\mathrm{Ni}}\,\text{\AA})]_{20}$
MLs, measured at room temperature ($300\,\mathrm{K}$) and a microwave
frequency of $9.8\,\mathrm{GHz}$. The resonance field is plotted as a
function of the angle $\theta$ of the applied magnetic field, measured
with respect to the in-plane direction.}
    \label{fig:FMRpolar}
\end{figure}

The resonance field ($H_{\mathrm{r}}$) was determined at each angle from the zero crossing of the derivative FMR spectrum when a single resonance was observed. For the perpendicular configuration ($\theta=90^\circ$), where multiple resonances were present, the spectra were integrated to obtain the absorption profiles. The field corresponding to the dominant absorption maximum was then assigned to the uniform-mode resonance, as illustrated in Fig.~\ref{fig:FMRint}. Fig.~\ref{fig:FMRpolar} shows the angular dependence of $H_{\mathrm{r}}$, where $\theta$ is measured between the applied magnetic field and the sample surface. The polar profiles exhibit twofold symmetry, with minimum resonance fields at $\theta=0^\circ$ and $180^\circ$ and maximum values at $\theta=90^\circ$ and $270^\circ$. Thus, a higher field is required to achieve resonance as the field direction changes from parallel to perpendicular to the film plane. This behaviour indicates an effective easy-plane magnetic anisotropy, consistent with previous findings for Si/Ni MLs~\cite{Singh2022SD} and other magnetic thin films~\cite{Jha2021}. The angular variation of $H_{\mathrm{r}}$ becomes more pronounced with increasing Ni thickness, mainly because the perpendicular resonance shifts towards higher fields, while the in-plane resonance fields remain relatively close. In contrast, the thinner Ni layers show a smaller difference between the two orientations. This is particularly evident for $t_{\mathrm{Ni}}=20$~\AA{}, which exhibits the least elongated polar profile and the weakest angular dependence among the samples.

\begin{figure}[!tbp]
    \centering
    \includegraphics[width=\linewidth]{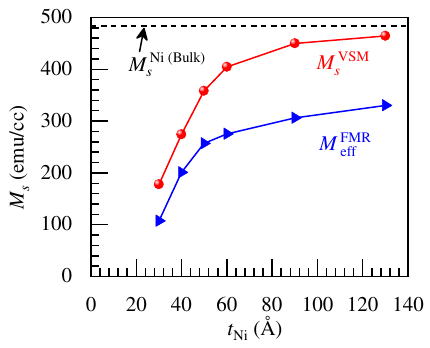}
    \caption{The plot of the saturation magnetization
($M_{\mathrm{s}}$, red solid circles) and effective magnetization
($M_{\mathrm{eff}}$, solid triangles) as a function of Ni layer thickness
($t_{\mathrm{Ni}}$) in $[\mathrm{Si}(40\,\text{\AA})/\mathrm{Ni}(t_{\mathrm{Ni}}\,\text{\AA})]_{20}$
MLs at ambient temperature, determined using VSM and FMR measurements, respectively. }
    \label{fig:meff}
\end{figure}

Apart from the resonance positions, the parallel spectra in Fig.~\ref{fig:fmr} exhibit unequal positive and negative derivative amplitudes. This departure from a symmetric line shape may contain a conductivity-related contribution associated with microwave skin effects~\cite{Dubiel2017EPJ}. The degree of asymmetry was quantified using the ratio $A/B$, where $A$ and $B$ denote the absolute amplitudes of the positive and negative portions of the derivative signal, respectively. An asymmetry ratio close to unity indicates a nearly symmetric spectrum. It is found that as $t_{\mathrm{Ni}}$ decreases, the ratio $A/B$ falls from 1.02 to 0.80, indicating an increasing departure from symmetry. This trend may be associated with the greater interface roughness at smaller Ni thicknesses, as established by the previously reported X-ray reflectivity analysis. A similar correlation was observed in discontinuous Si/Ni MLs, where the largest FMR asymmetry was associated with the highest surface roughness measured by atomic force microscopy~\cite{Singh2022SD}.

The resonance fields at $\theta = 0^\circ$ and $\theta = 90^\circ$,  were used to determine the effective magnetization, $M_{\mathrm{eff}}^{\mathrm{FMR}}$, by fitting the experimental data to Kittel's equations~\cite{Kittel2005}:
\begin{align}
\left(\frac{\omega}{\gamma}\right)_{\parallel}
&=
\sqrt{
H_{\parallel}
\left(
H_{\parallel}+4\pi M_{\mathrm{eff}}^{\mathrm{FMR}}
\right)
}
\label{eq:kittel_parallel}
\\
\left(\frac{\omega}{\gamma}\right)_{\perp}
&=
H_{\perp}
-4\pi M_{\mathrm{eff}}^{\mathrm{FMR}}
\label{eq:kittel_perpendicular}
\end{align}
where $\Meff$ is the effective magnetization and $\omega=2\pi f$ is the angular microwave frequency. The same excitation frequency applies to both measurement geometries. The gyromagnetic ratio is related to the spectroscopic splitting factor through $\gamma=g\mu_{\mathrm{B}}/\hbar$, where $\mu_{\mathrm{B}}$ is the Bohr magneton and $\hbar$ is the reduced Planck constant. Consistent units are used throughout the resonance analysis. For comparison, the reported bulk Ni value of $g$ is approximately 2.21~\cite{Cullity2011, Singh2022SD}. However, the extracted spectroscopic splitting factors lie between 2.14 and 2.26, which encompass the bulk Ni.

Fig.~\ref{fig:meff} compares the effective magnetization, $M_{\mathrm{eff}}^{\mathrm{FMR}}$, extracted from the uniform resonance mode using Eqs.~\ref{eq:kittel_parallel} and~\ref{eq:kittel_perpendicular}, with the saturation magnetization, $M_{\mathrm{s}}^{\mathrm{SVSM}}$, measured using VSM for the Si/Ni MLs. Both quantities rise sharply at smaller $t_{\mathrm{Ni}}$ and more gradually for thicker Ni layers, consistent with earlier reports~\cite{Singh2022SD, Singh2016cnf}. This trend may reflect the growth of Ni nanocrystallites and improved layer continuity~\cite{Singh2022SD}. With increasing $t_{\mathrm{Ni}}$, $M_{\mathrm{s}}^{\mathrm{VSM}}$ approaches the bulk Ni value of $484\,\mathrm{emu\,cm^{-3}}$ at $300\,\mathrm{K}$~\cite{Kittel2005}, indicated by the dashed horizontal line. Its reduction relative to bulk Ni may be associated with magnetically dead regions at the Si/Ni interfaces, characterized by an estimated thickness $\delta = 9\,\text{\AA}$, which reduces the effective magnetic thickness of the Ni layers.

For all samples, $M_{\mathrm{eff}}^{\mathrm{FMR}}$ remains lower than $M_{\mathrm{s}}^{\mathrm{VSM}}$, as shown in Fig.~\ref{fig:meff}.
However, the magnitude of this difference depends on the Ni layer
thickness, $t_{\mathrm{Ni}}$. At $t_{\mathrm{Ni}} = 30-40\,\text{\AA}$, the difference is relatively small, whereas it increases with increasing Ni thickness and becomes more pronounced for the thicker layers. Within the framework of the Kittel model, this discrepancy can be attributed to a net PMA contribution, which may originate from both interfacial and volume anisotropies~\cite{Singh2022SD}. The effective
magnetization extracted from FMR measurements reflects the combined
influence of the saturation magnetization and the perpendicular
anisotropy field, whereas the saturation magnetization obtained from
VSM measurements represent the magnetic moment per unit volume.
Despite the demagnetizing contribution, the perpendicular anisotropy field, $H_K$, can be expressed as \cite{Lassri2001TSF,Alayo2015JMMM, Gabor2019},

\begin{equation}
  H_\mathrm{K} = \frac{2K_{\perp}}{\mu_0 M_s^{\mathrm{VSM}}}
  = 4\pi\left(M_s^{\mathrm{VSM}} - M_{\mathrm{eff}}^{\mathrm{FMR}}\right).
  \label{eq:hk}
\end{equation}\\
where $K_{\perp}$ is the effective perpendicular anisotropy, including interfacial and volume contributions, and $M_s$ is the saturation magnetization. For a positive perpendicular anisotropy field $H_K$, the condition $M_{\mathrm{eff}}^{\mathrm{FMR}} < M_s^{\mathrm{VSM}}$ indicates
a net perpendicular anisotropy contribution. Furthermore, since demagnetization effects are intrinsic to thin-film magnetic systems and significantly influence their magnetic anisotropy, they must be considered when evaluating the effective magnetic anisotropy, $K_{\mathrm{eff}}$. In Gaussian cgs units, $K_{\mathrm{eff}}$ is given by
\begin{equation}
    K_{\mathrm{eff}} = K_{\perp} - 2\pi M_s^2,
    \label{eq:keff}
\end{equation}

By combining the above expression with Eq.~\ref{eq:hk}, $K_{\mathrm{eff}}$ is estimated, which can generally be expressed as the sum of the volume and interfacial contributions ~\cite{Lassri2001TSF, Lassri2004PB, Abid1999JMMM}. The volume anisotropy term, $K_V$, accounts for contributions arising from magnetocrystalline anisotropy and strain-induced anisotropy, whereas the interfacial anisotropy term, $K_S$, originates from the magnetic interactions at the interfaces of the ferromagnetic layer. Since the interfacial contribution is normalized by the ferromagnetic layer thickness, $t_{\mathrm{Ni}}$, the effective anisotropy energy density can be written as
\begin{equation}
    K_{\mathrm{eff}} = K_V + \frac{2K_S}{t_{\mathrm{Ni}}},
    \label{eq:keff}
\end{equation}
where $K_V$ is the volume anisotropy, and $K_S$ is the
interfacial anisotropy energy per unit area. This relation provides a convenient framework for distinguishing the volume and interfacial contributions to the effective magnetic anisotropy by analyzing its dependence on the ferromagnetic layer thickness. The expression above accounts for the respective contributions of the interfacial and bulk-like atoms to the overall MAE of the Ni layer, with their relative weights determined by the layer thickness $\tNi$~\cite{Johnson1996RPPa}. The coefficient 2 accompanying $\Ks/\tNi$ reflects the presence of two equivalent interfaces on either side of the Ni layer.

\begin{figure}[!tbp]
    \centering
    \includegraphics[width=\linewidth]{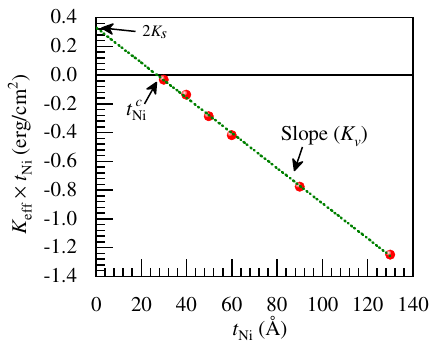}
    \caption{The effective anisotropy energy density multiplied by the Ni layer thickness ($K_{\mathrm{eff}}.t_{\mathrm{Ni}}$), is plotted as a function of
$t_{\mathrm{Ni}}$ for the
$[\mathrm{Si}(40\,\text{\AA})/\mathrm{Ni}(t_{\mathrm{Ni}}\,\text{\AA})]_{20}$
MLs at $300\,\mathrm{K}$. The experimental results are represented
by solid symbols, while the solid line denotes the linear fit obtained
using Eq.~\eqref{eq:keff}.}
    \label{fig:keff}
\end{figure}

Fig.~\ref{fig:keff} shows the Ni layer thickness dependence of $K_{\mathrm{eff}}.t_{\mathrm{Ni}}$, calculated using Eq.~\ref{eq:keff}, for the Si/Ni MLs at $300\,\mathrm{K}$. The product decreases approximately linearly from nearly zero at $t_{\mathrm{Ni}} = 27\,\text{\AA}$ to about $-1.25\,\mathrm{erg\,cm^{-2}}$ at $130\,\text{\AA}$. Its negative values indicate effective easy-plane anisotropy under the convention that positive $K_{\mathrm{eff}}$ favours perpendicular magnetization. The negative slope and positive extrapolated intercept reveal competing volume and interfacial contributions, favouring in-plane and perpendicular alignment, respectively. As the Ni layers become thicker, the relative interfacial contribution diminishes, strengthening the easy-plane character. The phenomenological model predicts a linear dependence, yielding an interfacial anisotropy constant of $\Ks = 0.17\,\mathrm{erg\,cm^{-2}}$ and a volume anisotropy contribution of $\Kv = -1.23\times10^{6}\,\mathrm{erg\,cm^{-3}}$. The corresponding crossover thickness, $t_{\mathrm{c}}$, at which the effective anisotropy changes sign, is determined by $t_{\mathrm{c}} = -2\Ks/\Kv$, giving $t_{\mathrm{c}}\approx27\,\text{\AA}$. This value should be interpreted as a model-based extrapolation rather than direct evidence of a reorientation transition, as the continuous layer approximation may become unreliable for the thinnest samples.  The extracted anisotropy constants are compared with those reported for discontinuous Si/Ni MLs, namely $\Ks = 0.11\,\mathrm{erg\,cm^{-2}}$ and $\Kv = -1.02\times10^{6}\,\mathrm{erg\,cm^{-3}}$~\cite{Singh2022SD}. The present study of MLs exhibit a larger $\Ks$ and $\Kv$ in magnitude, indicating a stronger perpendicular interfacial contribution and a weaker in-plane volume contribution. These differences may be associated with the thicker magnetically dead regions in the present samples, which suggest modified interfacial magnetic properties and a reduced effective magnetic thickness. The anisotropy constants are also comparable in magnitude to those reported for other Ni-based layered systems~\cite{Johnson1996RPPa,Chafai2011JMMM,denBroeder1991JMMM}. For Ni/Pd MLs, Chafai et al.~\cite{Chafai2011JMMM} reported $\Ks = 0.03\,\mathrm{erg\,cm^{-2}}$ and $\Kv = -1.40\times10^{6}\,\mathrm{erg\,cm^{-3}}$, whereas den Broeder et al.~\cite{denBroeder1991JMMM} obtained a negligible interfacial term and volume contributions of $-1.25\times10^{6}$ and $-1.73\times10^{6}\,\mathrm{erg\,cm^{-3}}$ for vapour-deposited and sputtered samples, respectively. The latter difference was associated with deposition-induced tensile and compressive stresses. For epitaxial Ni/Co/Ni sandwiches grown by molecular beam epitaxy at $300\,\mathrm{K}$, Johnson et al.~\cite{Johnson1992PRL} reported $\Ks = 0.19$--$0.42\,\mathrm{erg\,cm^{-2}}$ and $\Kv = -4.50\times10^{6}$ to $-0.80\times10^{6}\,\mathrm{erg\,cm^{-3}}$ using magneto-optical Kerr measurements. Their larger interfacial anisotropy compared with the present Si/Ni MLs may reflect differences in crystalline order, interface composition, and strain~\cite{Johnson1996RPPa}.

\section{Conclusion}\label{sec13}

The room-temperature magnetic properties and magnetic anisotropy of continuous Si(40~\AA)/Ni($t_{\mathrm{Ni}}$~\AA)$_{20}$ MLs ($t_{\mathrm{Ni}} = 20$--$130$~\AA) deposited on Si(111) substrates were systematically investigated. VSM measurements reveal an increase in saturation magnetization toward the bulk Ni value ($484~\mathrm{emu\,cm^{-3}}$) and a reduction in coercivity with increasing Ni thickness. Dead-layer analysis yields an intrinsic magnetization of $M_0 \approx 500~\mathrm{emu\,cm^{-3}}$ and an interfacial magnetic dead-layer thickness of $\delta \approx 9$~\AA, accounting for the enhanced superparamagnetic contribution in thinner Ni layers.

Angle-dependent FMR measurements reveal twofold symmetry and predominantly easy-plane magnetic anisotropy. The resonance spectra further exhibit uniform and non-uniform spin-wave modes, while increasing line-shape asymmetry at reduced Ni thicknesses is consistent with enhanced interfacial roughness. The effective magnetization extracted from FMR remains lower than the static saturation magnetization, indicating a perpendicular anisotropy contribution. Quantitative analysis yields an interfacial anisotropy constant of $K_S = +0.17~\mathrm{erg\,cm^{-2}}$, favoring perpendicular magnetization, and a volume anisotropy constant of $K_V = -1.23 \times 10^6~\mathrm{erg\,cm^{-3}}$, favoring in-plane alignment, with an extrapolated crossover thickness of $t_c \approx 27$~\AA. Compared with previously reported discontinuous Si/Ni MLs, these continuous structures exhibit enhanced interfacial anisotropy and reduced in-plane volume anisotropy. These findings highlight the crucial role of Ni thickness and interface quality in controlling magnetic anisotropy and magnetization dynamics, providing valuable insights for designing Si-compatible spintronic heterostructures.

\section*{Acknowledgements}

The authors acknowledge financial support from the Science and Engineering Research Board (SERB), Department of Science and Technology (DST), Government of India, under Grant No.~EEQ/2022/000354. The facilities provided by the Central Instrumentation Facility (CIF), IIT Gandhinagar, for material characterization are gratefully acknowledged. The authors also acknowledge support from the DST-FIST program under Grant No.~SR/FST/PS-I/2020/147 (C), which enabled magnetization measurements using the Physical Property Measurement System (PPMS) at the Cryogen-Free Measurement Facility, IIT Gandhinagar.

\section*{Declarations}

The authors declare that they have no known competing financial interests or personal relationships that could have appeared to influence the work reported in this paper.

\bibliographystyle{unsrtnat}
\bibliography{sn-bibliography}

\end{document}